To be published in: Handbook of Exoplanets, 2nd Edition, Hans Deeg and Juan Antonio Belmonte (Eds. in Chief), Springer International Publishing AG, part of Springer Nature

# Transit Photometry as an Exoplanet Discovery Method

Hans J. Deeg[1, 2] and Roi Alonso[1, 2]

(1)Instituto de Astrofísica de Canarias, La Laguna, Tenerife, Spain

(2)Departamento de Astrofísica, Universidad de La Laguna, La Laguna, Tenerife, Spain

## Abstract

Photometry with the transit method has arguably been the most successful exoplanet discovery method to date. A short overview about the rise of that method to its present status is given. The method's strength is the rich set of parameters that can be obtained from transiting planets, in particular in combination with radial velocity observations; the basic principles of these parameters are given, with explicit formulations for the transit detection probability and the times of transit epochs in comparison to radial velocity epochs. The transit method has however also drawbacks, which are the low probability of properly aligned planet systems and the presence of astrophysical phenomena that may mimic transits and give rise to false detection positives. In the second part, we outline the main factors that determine the design of transit surveys, such as the size of the survey sample, the temporal coverage, the photometric precision, the sample brightness and the methods to extract transit events from observed light curves. Lastly, an overview over past, current, and future transit surveys is given. For these surveys we indicate their basic instrument configuration and their planet catch, including the ranges of planet sizes and stellar magnitudes that were encountered. Current and future transit detection experiments concentrate primarily on bright or special targets, and we expect that the transit method remains a principal driver of exoplanet science, through new discoveries to be made and through the development of new generations of instruments.

## Introduction

Since the discovery of the first transiting exoplanet, HD 209458b (Henry et al. *2000*; Charbonneau et al. *2000*), the transit method has become the most successful detection method, surpassing the combined detection counts of all other methods (see Fig. 1) and giving rise to the most thoroughly characterized exoplanets at present.

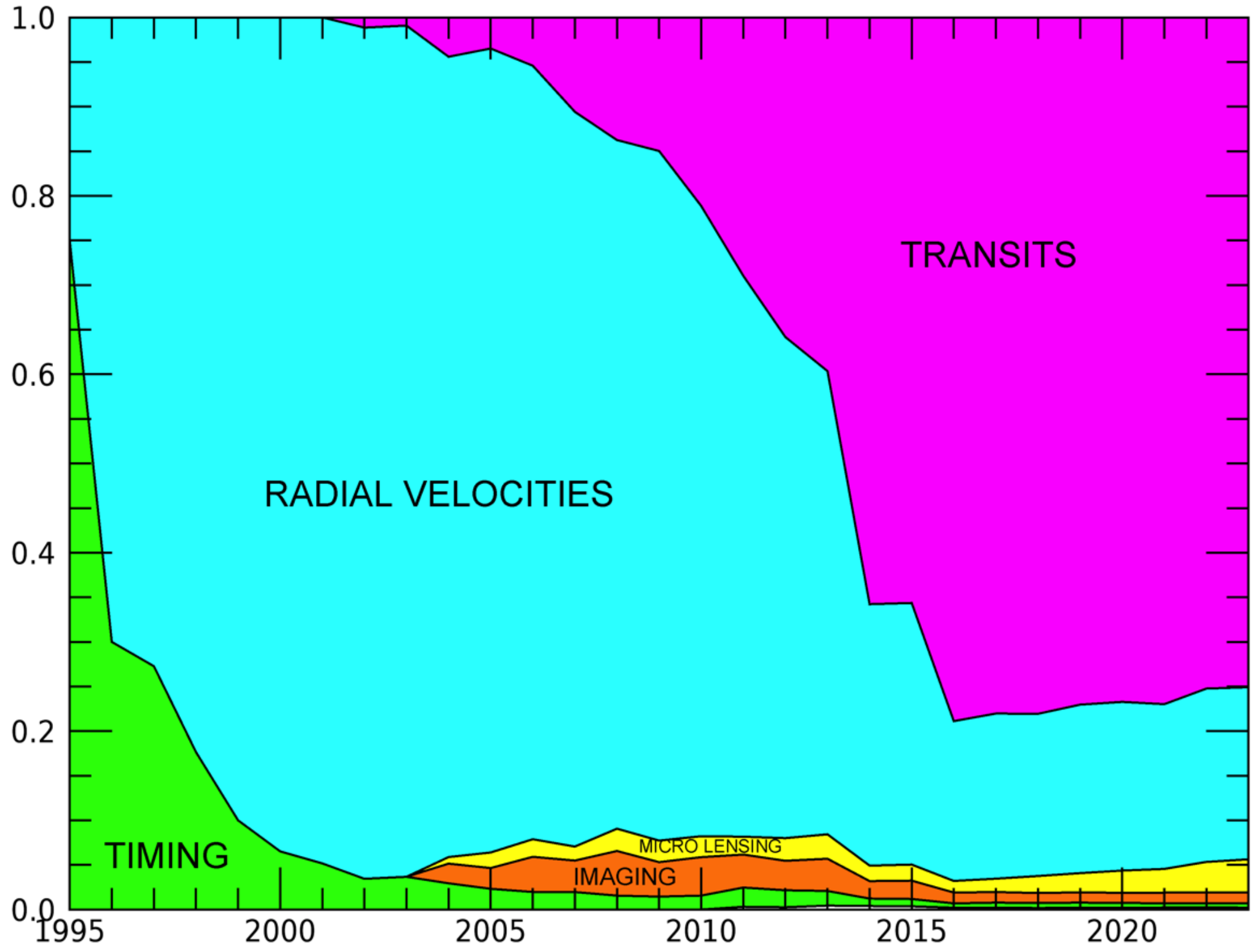

**Fig. 1**

The fractions by which various detection methods contributed to the accumulated sample of known planets are shown, for years since 1995. At the end of 1995, only four planets were known, three from pulsar timing and one from radial velocities. Between 1996 and 2013, the sample of known planets was dominated by those discovered with radial velocities, while since 2016, planets discovered by transits have maintained a fraction of about 75%, with a total of 5535 planets (October 2023) discovered by all methods combined. Based on data from the NASA Exoplanet Archive in and using its classification by discovery methods. 'Timing' includes planets found by pulsar timing, eclipse timing, or transit timing. Other detection methods (astrometry, orbital brightness variations) generate only a very small contribution that is barely visible at the bottom of the graph, for years following 2010

The detection of planetary transits is among the oldest planet detection methods; together with the radial velocity (RV) method, it was proposed in *1952* in a brief paper by Otto Struve (see also Chap. 3, "Prehistory of Transit Searches"). The early years of exoplanet discoveries were however dominated by planets found by RVs, and prior to the 1999 discovery of transits on HD 209458, the transit method was not considered overly promising by the community at large. For example, a *1996* (Elachi et al.) NASA Road Map for the Exploration of Neighboring Planetary Systems (ExNPS) revises in some detail the potential of RV, astrometry, and microlensing detections, with a recommended focusing onto space interferometry, while transits were considered only cursory. Consequently, activities to advance transit detections were rather limited; most notable are early proposals for a space-based transit search by Borucki, Koch, and collaborators (Borucki et al. *1985*; Koch et al. *1996*; Borucki et al. *1997*; see also Chap. 56, "Space Missions for Exoplanet Science: Kepler/K2") and the TEP project, a search for transiting planets around the eclipsing binary CM Draconis that had started in 1994 (Deeg et al. *1998*; Doyle et al. *2000*; see also Chap. 5, "The Way to Circumbinary Planets"). The discovery of the first transiting planets, with some of them like HD 209458b already known from RV detections, quickly led to intense activity to more deeply characterize them, mainly from multicolor photometry (e.g., Jha et al. *2000*; Deeg et al. *2001*) or from spectroscopy during transits (e.g., Queloz et al. *2000*; Charbonneau et al. *2002*; Snellen *2004*); to provide the community with efficient transit fitting routines (e.g., Mandel and Agol *2002*; for more see below), or to extract the most useful set of physical parameters from transit light curves (Seager and Mallén-Ornelas *2003*).

The first detections of transits also provided a strong motivation toward the setup of dedicated transit searches, which soon led to the first planet discoveries by that method, namely, OGLE-TR-56b (Konacki et al. *2003*) and further planets by the OGLE-III survey, followed by TrES-1 (Alonso et al. *2004a*), which was the first transit discovery on a bright host star. Transits were therefore established as a valid method to find new planets (see also Chap. 4, "Discovery of the First Transiting Planets").

Central to the method's acceptance was also the fact that planets discovered by transits across bright host stars permit the extraction of a wealth of information from further observations. Transiting planets orbiting bright host stars, such as HD 209458b, HD 189733 (Bouchy et al. *2005*), WASP-33b (Christian et al. *2006*; Collier Cameron et al. *2010*), or the terrestrial planet 55 Cnc e (McArthur et al. *2004*; Winn et al. *2011*), are presently the planets about which we have the most detailed knowledge; an updated list of the well-studied transiting planets is provided by TEPCat (Southworth 2011; https://www.astro.keele.ac.uk/jkt/tepcat/). Besides RV observations for the mass and orbit determinations, further characterization may advance with the following techniques: transit photometry with increased precision or in different wavelengths, transit photometry to derive transit timing variations (TTVs), and spectroscopic observations during transits (transmission spectroscopy, line-profile tomography of exoplanet transits, Rossiter-McLaughlin effect). Furthermore, the presence of transits – strictly speaking primary transits of a planet in front of its central star – usually implies the presence of secondary eclipses or occultations, when a planet disappears behind its central star. These eclipses, as well as phase curves of a planet's brightness in dependence of its orbital position, might be observable as well (see corresponding chapters for these techniques).

Given the modest instrumental requirements to perform such transit searches on bright star samples – both HD209458b's transits and the planet TrES-1 were found with a telescope of only 10 cm diameter – the first years of the twenty-first century saw numerous teams attempting to start their own transit surveys. Also, the two space-based surveys that were launched a few years later, CoRoT and Kepler, are unlikely to have received the necessary approvals without the prior ground-based discovery of transiting planets (see Chap. 54, "Space Missions for Exoplanet Research: Overview and Introduction" and chapters on the individual missions). The enthusiasm for transit search projects at that time is well represented by a paper by Horne (*2003*) which lists 23 transit surveys that were being prepared or already operating. Its title "Hot Jupiters Galore" also typifies the expectation that significant numbers of transiting exoplanets will be found in the near future: Summing all 23 surveys, Horne predicted a rate of 191 planet detections per month! In reality, advances were much slower, with none of these surveys reaching the predicted productivity. By the end of 2007, before the first space-based discoveries from the CoRoT mission (Barge et al. *2008*; Alonso et al. *2008*), only 27 planets had been found through transit searches. This slower advance can be traced to two issues that revealed themselves only during the course of the first surveys: The amount of survey time required under real conditions was higher than expected, and the presence of red noises degraded sensitivity to transit-like events (see later in this chapter). Once these issues got understood and accounted for, some of these ground-based surveys became very productive, and both the SuperWASP and the HAT surveys (HATnet and HATSouth) have detected over 140 planets to date (see Table 1 for details). The next major advances based on the transit method arrived with the launch of the space missions CoRoT in 2006 and Kepler in 2009. These led to the discoveries of transiting terrestrial -sized planets (CoRoT-7b by Léger et al. 2009, Kepler-10b by Batalha et al. 2011) to planets in the temperate regime (CoRoT-9b, Deeg et al. 2010), to transiting multi-planet systems (Lissauer et al. 2011), and to a huge amount of transiting planets that permit a deeper analysis of planet abundances in a very large part of the radius – period (or Teff) parameter space (see Chap. 96, "Planet Occurrence: Doppler and Transit Surveys"). Also, the first candidates for Exo-Moons have been claimed from transits in Kepler lightcurves (Teachey, Kipping & Schmitt 2018, Kipping et al. 2022, with a refutation by Heller & Hippke 2023). The next major step came in 2018 with the launch of the TESS (Transiting Exoplanets Survey Satellite) mission, (Ricker et al. 2015), which is scanning nearly the entire sky for the presence of transiting planets, and which will lead to a nearly complete discovery of all transiting planets with periods of less than a month and with suitable parameters (mainly transit depth and system brightness) for discovery by that mission. In the following, an introduction is given on the methodology of the transit detection and its surveys, as well as an overview about the principal projects that implement these surveys.

# Fundamentals of the Transit Method

A schematic view of a transit event is given in Fig. 2, where the bottom part represents the observed flux of the system. As the planet passes in front of the star, its flux diminishes by a fractional denoted as $\Delta F$. Under the assumptions of negligible flux from the planet and of spherical shapes of the star and planet, $\Delta F$ is given by the ratio of the areas of the planet and the star:

$$\Delta F \approx \left(\frac{R_p}{R_s}\right)^2 = k^2 \tag{1}$$

where $Rp$ is the radius of the planet, $Rs$ the radius of the star, and $k$ is the radius ratio. The total duration of the transit event is represented as $tT$, and the time of totality, in which the entire planet disk is in front of the stellar disk (the time between *second* and *third* contacts, using eclipse terminology), is given by $tF$, during which the light curve is relatively flat. Using basic geometry, the work by Seager and Mallén-Ornelas (2003) derived analytic expressions that relate these observables to the orbital parameters. In particular, the impact parameter $b$, defined as the minimal projected distance to the center of the stellar disk during the transit, can be expressed as:

$$b \equiv \frac{a}{R_s}\cos i = \left\{\frac{(1-k)^2 - [\sin^2(t_F\pi/P)/\sin^2(t_T\pi/P)](1+k)^2}{\cos^2(t_F\pi/P)/\cos^2(t_T\pi/P)}\right\}^{1/2} \tag{2}$$

where $a$ is the orbital semimajor axis, $i$ the orbital inclination, and $P$ the orbital period. A commonly used quantity that can be obtained from photometric data alone is the so-called scale of the system or the ratio between the semimajor axis and the radius of the star:

$$\frac{a}{R_s} = \sqrt{\frac{(1+k)^2 - b^2\cos^2(t_T\pi/P))}{\sin^2(t_T\pi/P)}} \tag{3}$$

which, using Kepler laws of motion and making the reasonable approximations of the mass of the planet being much smaller than the mass of its host star, and assuming a spherical shape for the star, can be transformed into a measurement of the mean stellar density:

$$\rho_s = \frac{3\pi}{GP^2}\left(\frac{a}{R_s}\right)^3 \quad (4)$$

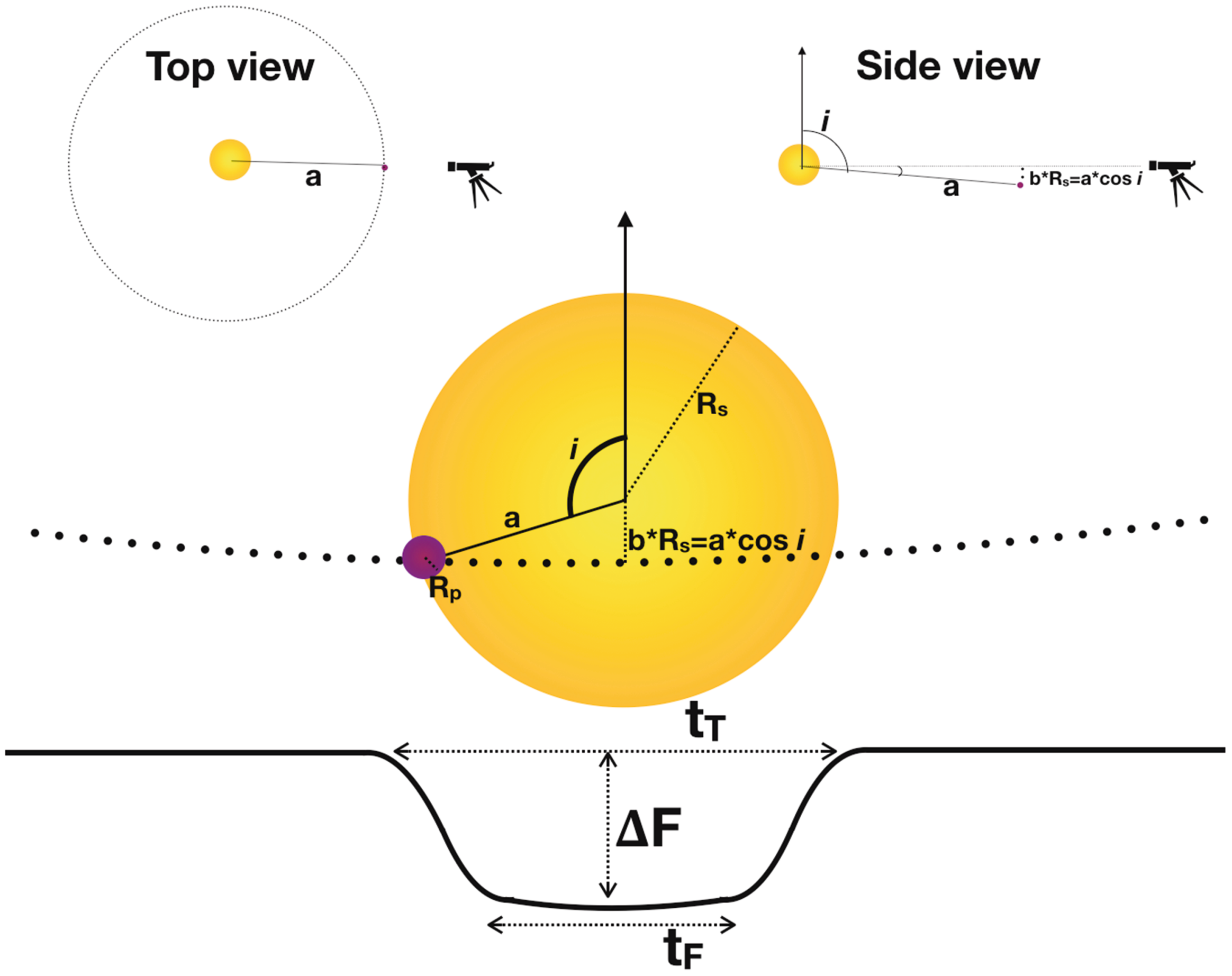

**Fig. 2**

Outline of the transit of an exoplanet, with the main quantities used to describe the orbital configuration, from the observables given in the lower solid curve (the observed light curve) to the model representations from the observer's point of view (central panel) or other viewpoints (top panels). Note that in the central panel, projected views of *a* and *i* are drawn

Consistency between this measurement and a stellar density estimated by other means (through spectroscopy, mass-radius relations, or asteroseismology) has often been used as a way to prioritize the best transit candidates from a survey (Seager and Mallén-Ornelas 2003; Sozetti et al. 2007; Tingley et al. 2011; Kipping et al. 2014). In the previous equations, we have assumed circular orbits for simplicity; a derivation of equivalent equations including the eccentricity terms can be found in Tingley et al. (2011, with a correction in Eq. 15 of Parviainen et al. 2013).

While the previous expressions allow quick estimates of the major parameters of an observed transit, more sophisticated methods using the formalisms of Mandel and Agol (2002) or Giménez (2006) are commonly used for their more precise derivation. This is in part due to a subtle effect visible in Fig. 2: the limb-darkening of the star that manifests itself as a nonuniform brightness of the stellar disk, which is described in detail in Chap. 67, "Stellar Limb Darkening's Effects on Exoplanet Characterization". The limb-darkening of the star makes it challenging to determine the moments of *second* and *contact* of the transit and to precisely measure $\Delta F$. For more detailed introductions into the parameters that can be measured from transits, we refer to Winn (2010), Haswell (2010) and for an analytical treatment relating the transit depth $k$ with the stellar limb-darkening, we refer to Heller (2019).

## Detection Probability

The probability that a randomly placed observer may observe a planet in transit is given by Seagroves et al. (2003), Winn (2010):

$$p_{tra} = \left(\frac{R_s \pm R_p}{a}\right)\left(\frac{1 + e\sin\omega}{1 - e^2}\right) \qquad (5)$$

where the + sign is used to include grazing transits and the − sign refers to the probability of full transits that have *second* and *third* contacts; e is the planet's orbital eccentric and ω is the argument of periastron of the stellar orbit (induced by the planet; for the planet's orbit it is $\omega_p = \omega \pm \pi/2$). From Eq. (5), the probability to observe transits of a planet of known eccentricity e but with unknown (uniformly distributed) ω may be derived as:

$$p_{tra} = \left(\frac{R_s \pm R_p}{a}\right)\left(\frac{1}{1 - e^2}\right) \qquad (6)$$

The detection probability of planets increases therefore with their eccentricity, albeit this effect is small for moderate for moderate eccentricities (e.g. for e = 0.5, $p_{tra}$ is elevated by 1.33 against circular orbits), but for e=0.8, this probability increases 2.8 times over the circular case.

For a typical Hot Jupiter with a semimajor axis of 0.05 AU, the detection probability is on the order of 10%, while for an Earth-like planet at 1 AU from a solar-like star, it goes down to 0.5%. These relatively low probabilities are the main handicap of the transit method, since the majority of existing planet systems will not display transits.

## Transit Epochs versus Radial Velocity Epochs

The relation between the epochs (time of reference) of radial velocity (RV) curves – be they from observed data or from modelling – and the epochs of transits (time of mid-transit) is a topic prone to confusion, and hence is described here in detail.

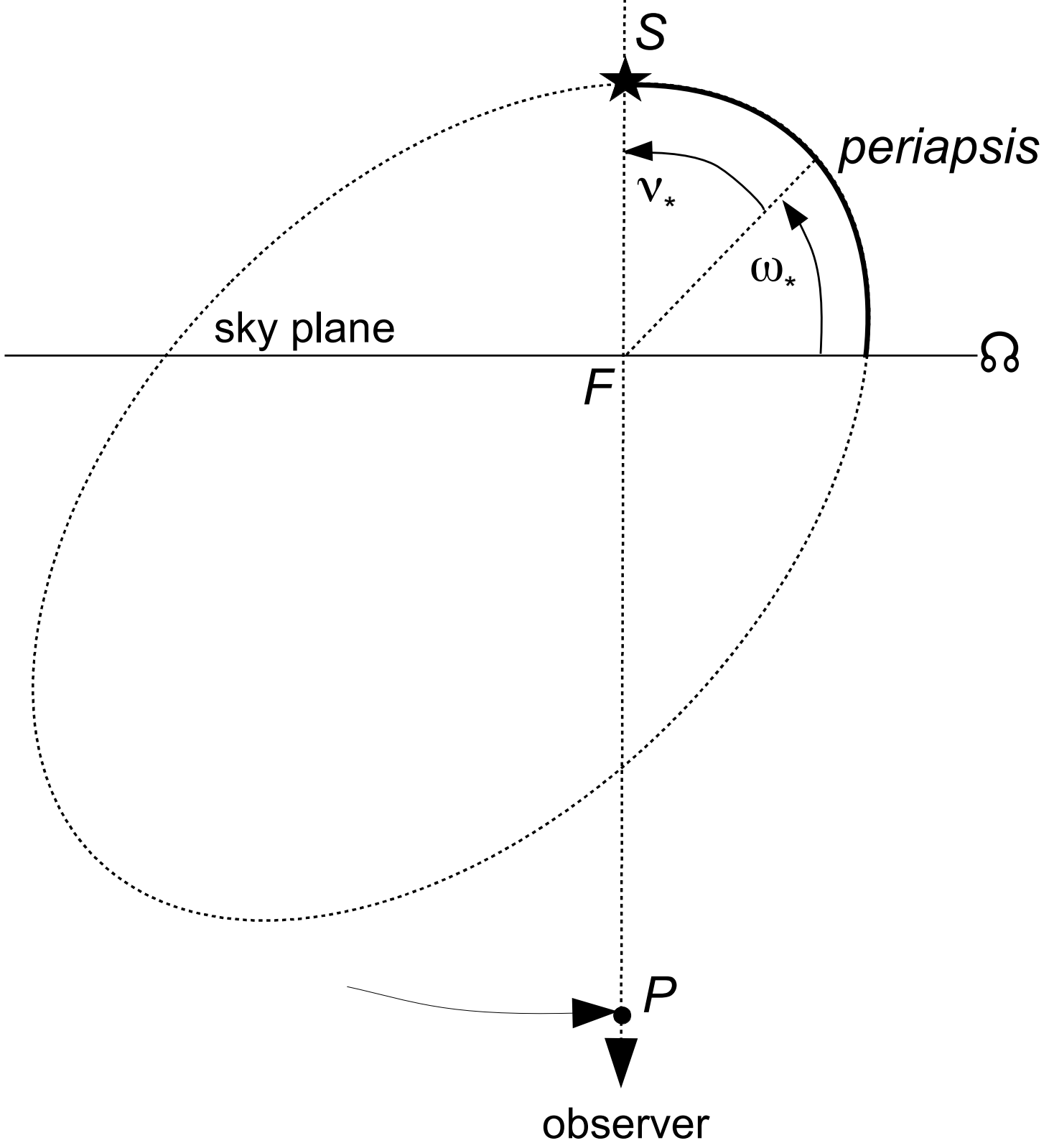

**Fig. 3**
Keplerian orbit of a star-planet system. F is the focal point of the star's (S) and planet's (P) elliptical orbits; it corresponds to the barycenter of the star-planet system. The sky plane is perpendicular to the line of sight from the observer and goes through the focal point. ω* is the star's angle of periapsis, which is described as the angle along the orbital trajectory, from ☊ (ascending node, where the object crosses the sky plane moving or 'ascending' away from the observer) to the periapsis. $\nu^*$ is the true anomaly of the star; it is an angle from the position at periapsis to the object's position at a given time. Only the star's orbit is shown in detail; the planet's orbit is an ellipse of identical shape, but larger and rotated by 180° around F

For the appearance of a transit at a time $T_c$ (transit center time), two conditions must be fulfilled: For one, the inclination of the planet's orbital plane needs to be sufficiently close to 90° so that transits are visible. This condition is of no concern in the following discussion, since its conclusions are not affected by the orbital inclination. The second condition is the passing of the star behind the planet-star barycenter, which happens at the same moment when the planet passes in front of it. From Fig **3**, it is obvious that this occurs when $\omega^* + \nu^* = 90°$ (for symbol definitions, see the figure caption), where we note that $\nu^*$ is a function of time. In most cases, $\omega^*$ is however larger than 90°; the transit epochs $T_c$ are hence given more generally by:

$$\nu^*(T_c - T_{RV}) = 90° \pm (n \times 360°) - \omega^* \ , \qquad (7)$$

where n is any integer and $T_{RV}$ is the epoch of the Keplerian orbit of the star, defined as the epoch when the star passes through its periapsis (and hence, $\nu^*(T_{RV}) = 0$ ).

Obtaining the difference between the transit and RV epochs, $\Delta T = T_c - T_{RV}$, requires an inversion of $\nu^*(\Delta T)$ as a function on the mean anomaly $M(\Delta T)$. We note that M is identical to the commonly used

RV phases ranging from 0 to 1, but ranging from 0 to 360°, or 0 to $2\pi$. The inversion needs to be performed numerically and is well described in the literature; it commonly uses the eccentric anomaly E as an intermediate value.

If the angle of periastron of the orbit of the planet, $\omega_p$, is used as an orbital parameter instead of $\omega^*$, we note that the true anomalies of planet and star are always the same, e.g. $\nu^* = \nu_p$, whereas the angles of periapsis are 180° apart, e.g. $\omega_p = \omega^* \pm 180°$. Eq. 7 is therefore modified to:

$$\nu_p(T_c - T_{RV}) = 270° \pm (n \times 360°) - \omega_p \tag{8}$$

Care has to be taken that an RV epoch from the literature corresponds really to the time of periapsis $T_{RV}$. While this is the mostly the case, $T_{RV}$ might be poorly defined for orbits of low eccentricity (because the uncertainty of $\omega^*$ might be very large, while in the circular orbits, $\omega^*$ can take any value); the usage of alternative parameters for the RV epochs might therefore be motivated. Examples for other kinds of RV epochs are the time of maximum RVs (e. g. in the famous paper by Mayor & Queloz 1995), or the epoch when RVs cross from positive to negative values. We note that for circular orbits, such an RV epoch corresponds to the transit epoch. Also common is the quoting of the Mean Anomaly M of the first RV data point, especially for low-eccentricity or circular orbits (where $M^* = \nu^*$). There, M=0 should correspond to the crossing of the sky-plane (e.g. $\omega^*$ =0° and $\nu^*$ = 0° in Fig. 3), but sometimes M=0 corresponds to $\omega^*$ =90°, $\nu^*$ = 0° , the epoch the transit. In all instances, for a correct modelling of a planet's orbit, a careful review of a quoted RV epoch is required in order to relate it to the epoch of transits. Some examples for Keplerian RV models with the epoch of transits are shown in Fig. 4

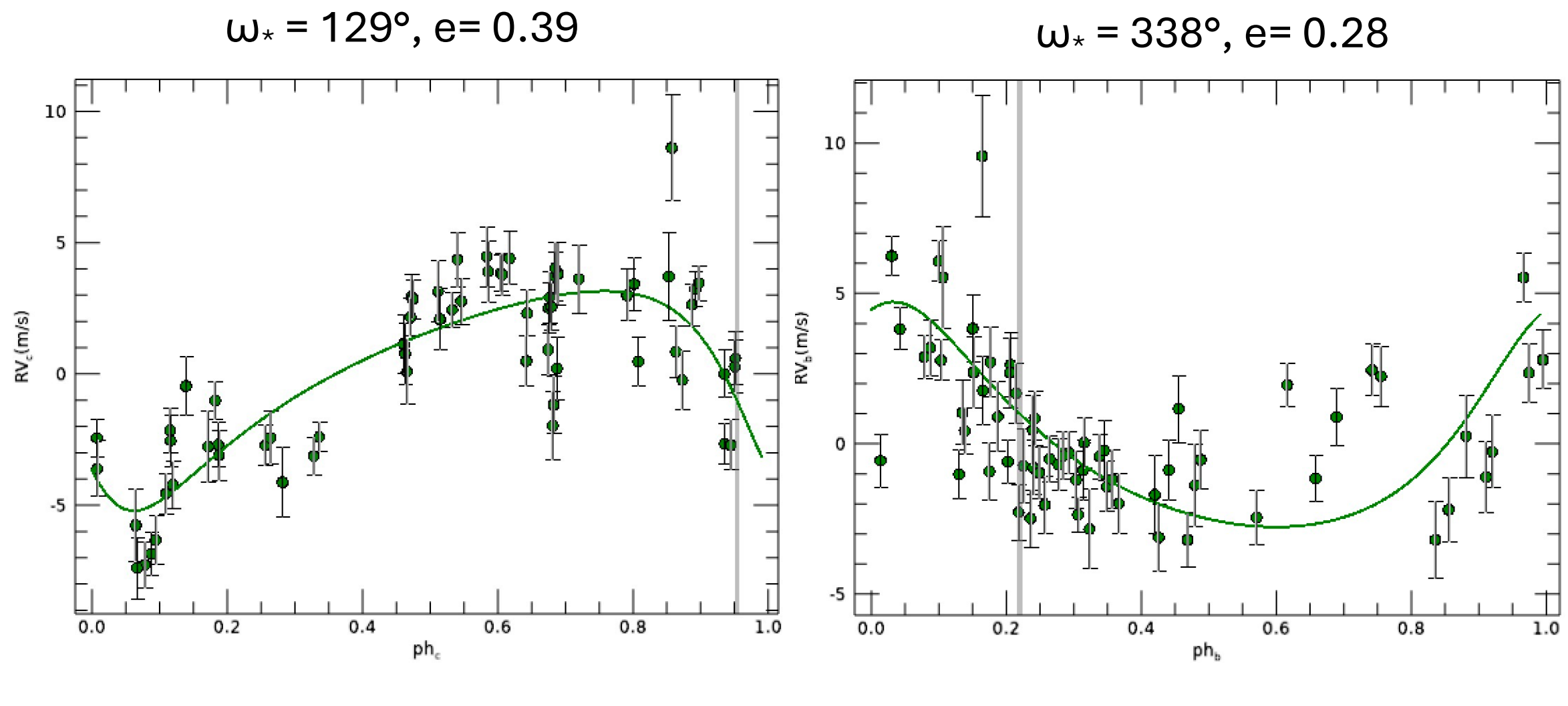

**Fig. 4**
Two examples of stellar radial velocities and Keplerian RV models (with the indicated orbital parameters), due to the presence of planet candidates, and the moments of transits (shown by the vertical grey bar). The horizontal axes give the RV phases, which are identical to the Mean Anomaly, but on a scale of 0 to 1 (instead of 0 to 360° or 0 to $2\pi$), with ph=0 or M=0 being the epoch when the star passes through periapsis; the vertical axes indicate relative radial velocities (against an RV average of zero). Here, like in all cases except for extreme eccentricities, transits occur close to the moment when RVs turn from positive to negative values. The models correspond to preliminary planet candidates in the TOI-1416 planet system, published in final form in Deeg et al. (2023)

## False Positives

A transit-shaped event in a light curve is not always caused by a transiting planet, as there are a number of astrophysical configurations that can lead to similar signatures. These are the so-called false positives in transit searches, which have been a nuisance of transit surveys since their early days. One example of a false positive would be a stellar eclipsing binary that is so close to a brighter single star that the light of both objects falls within the same photometric aperture of a detector: the deep eclipses of the eclipsing binary are diluted due to the flux of the brighter star, resulting in a light curve with a shallower eclipse, with a shape very similar to a transiting planet. More complete descriptions of the types of false positives that affect transit searches, and their expected frequencies, can be found in Brown (*2003*), Alonso et al. (*2004b*), Almenara et al. (*2009*), and Santerne et al. (*2013*).

To detect false positives and to confirm the planetary nature of a list of candidates provided by a transit survey, a series of follow-up observations are required (e.g., Latham *2003*; Alonso et al. *2004b*; Latham *2007*, *2008*; Deeg et al. *2009*; Moutou et al. *2013*; Günther et al. *2017*), which apply to both ground-based or space-based surveys. Traditionally, the *confirmation* that a transit signal is caused by a planet takes place when its mass is measured with high-precision RV measurements. In some cases, particularly with planets orbiting faint host stars, or for the confirmation of the smallest planets, the achievable RV precision is insufficient to measure the planet's mass. As these cases are of high interest, for example, planets with similar sizes as the Earth orbiting inside the habitable zone of its host star, statistical techniques have been developed to estimate the probability of the observed signals being due to planets relative to every other source of false positives we know of. In this case, the planets are known as *validated*. Current validation procedures use the fact that astrophysical false-positive scenarios have very low probabilities when several transiting signals are seen on the same star (Lissauer et al. *2012*), or they use all the available information (observables and knowledge of the galactic population and stellar evolution) to compare the probabilities of a signal being due to a transiting planet vs. anything else. A few examples of validation studies are Torres et al. (*2011*), Morton (*2012*), Lissauer et al. (*2014*), Rowe et al. (*2014*) , Díaz et al. (*2014*), Santerne et al. (2015), Torres et al. (*2015*), Morton et al. (*2016*), Torres et al. (*2017*) and Giacalone & Dressing (2020), some of which use these current state-of-the-art validation procedures: BLENDER, VESPA, triceratops or PASTIS, described in these references.

Finally, some false positives may be due to artifacts from red noise or other instrumental effects, even in the most precise surveys to date (e.g., Coughlin et al. *2014*). In a few cases, planets that were previously validated have been disproved after an independent analysis (Cabrera et al. *2017*; Shporer et al. *2017*), which should generate some caution about the use of results from validations, which are statistical by design.

# Transit Surveys: Factors Affecting their Performance

The task of surveying a stellar sample for the presence of transiting planets must overcome the inherent inefficiencies of the transit method: The planets need to be aligned correctly (see previous section), and the observations must be made when transits occur. The expected abundances of the desired planet catch must be taken into account, and their transits need to be detectable with sufficient photometric precision. Furthermore, transit-like events (false positives) may arise from other astrophysical as well as instrumental sources, and means to identify them need to be provided. The success of a transit detection experiment must take these factors into account, which are discussed in the following:

*Sample size*

The probability $p_{tra}$ for transits to occur in a given random-oriented system is between a few percent for Hot Jupiters and less than 0.1% for cool giant planets. In order to achieve a reasonable probability that $N$ transiting system will be found in a given stellar field, the number of surveyed stars (that is, stars for which light curves with sufficient precision for transit detection are obtained) should be at least $N_{\mathrm{survey}} \approx N/(p_{\mathrm{tra}} f)$, where $f$ is the fractional abundance of the detectable planet population in the stellar sample. For surveys of Hot Jupiters, with $f \approx 1\%$ of main-sequence (MS) stars (Wright et al. 2012; Mayor et al. 2011) , this leads to minimum samples of 2000 MS stars to expect a single transit discovery. Given that most stars in the bright samples of small-telescope surveys are not on the MS, sample sizes of 5000–10000 targets are however more appropriate. Survey fields that provide sufficient numbers of suitable stars, by brightness and by desired stellar type, need therefore be defined. The size of the sample is then given by the size of the field of view (*fov*) and by the spatial density of suitable target stars, which depends on the precision of the detector (primarily depending on the telescope aperture) and on the location of the stellar fields. Also, in most surveys, sample size is increased through successive observations of different fields.

*Temporal coverage*

At any given moment, the probability for the observation of a transit of a correctly aligned system is $p \approx tT/P$, where $tT$ is the duration of a transit and $P$ the orbital period. This probability goes from 5–8% for Ultra-short periodic planets over 2–3% for typical Hot Jupiter systems to 0.15% for an Earth-Sun alike. For an estimation of the number of transits for a given sample at a given time, we need to multiply this probability and the probability for correct alignment with the abundance of detectable planets. To determine a planet's period, of course at least two transits need to be observed. The requirement to observe three transit-like events that are periodic has however been habitual in ground-based observations, which are prone to produce transit-like events from meteorologic and other non-astronomic causes. Furthermore, for an increased S/N of transit detections, especially toward the detection of smaller planets, as well as toward a more precise derivation of physical parameters, the rule is "the more transits, the better." Continuous observational coverage is the most time-efficient way to achieve the observation of a minimum number of transits (e.g., $N_{\mathrm{tr,min}} \geq 3$) for a given system. However, only space missions are able to observe nearly continuously over a timescale of weeks, which is the only way to ascertain that transiting planets above some size threshold and below some maximum period are being detected with near certainty. Ground-based surveys, with their interruptions from the day/night cycle and from meteorological incidences, can only seek reasonable *probabilities* (but no certainty) to catch a desired number of transits from a given planet. Several instruments have also been installed at high geographic latitudes in order to take advantage of the long polar winter nights, especially on the Antarctic plateau (e.g. ASTEP on Dome C, Crouzet et al. 2018; the Chinese Antarctic Survey Telescopes on Dome A, Yang et al. 2023) but to date with a limited return for transit detections. The principal factor that determines the number of observed transits in a given discontinuous light curve is a planet's orbital phase (taken at some reference time, such as the beginning of observations) or its epoch (the time when one of its transits occurs); both are of course unknown prior to a planet's discovery. An example of the effect of phase on the number of observed transits in discontinuous data is shown in Fig. 5. As a rough rule, in order to achieve reasonable detection probabilities (e.g., ≥ 70%) for typical Hot Jupiters ($P$ = 3−4 d) with a requirement of 3 observed transits, ground-based surveys should cover a stellar field for at least 300 h.

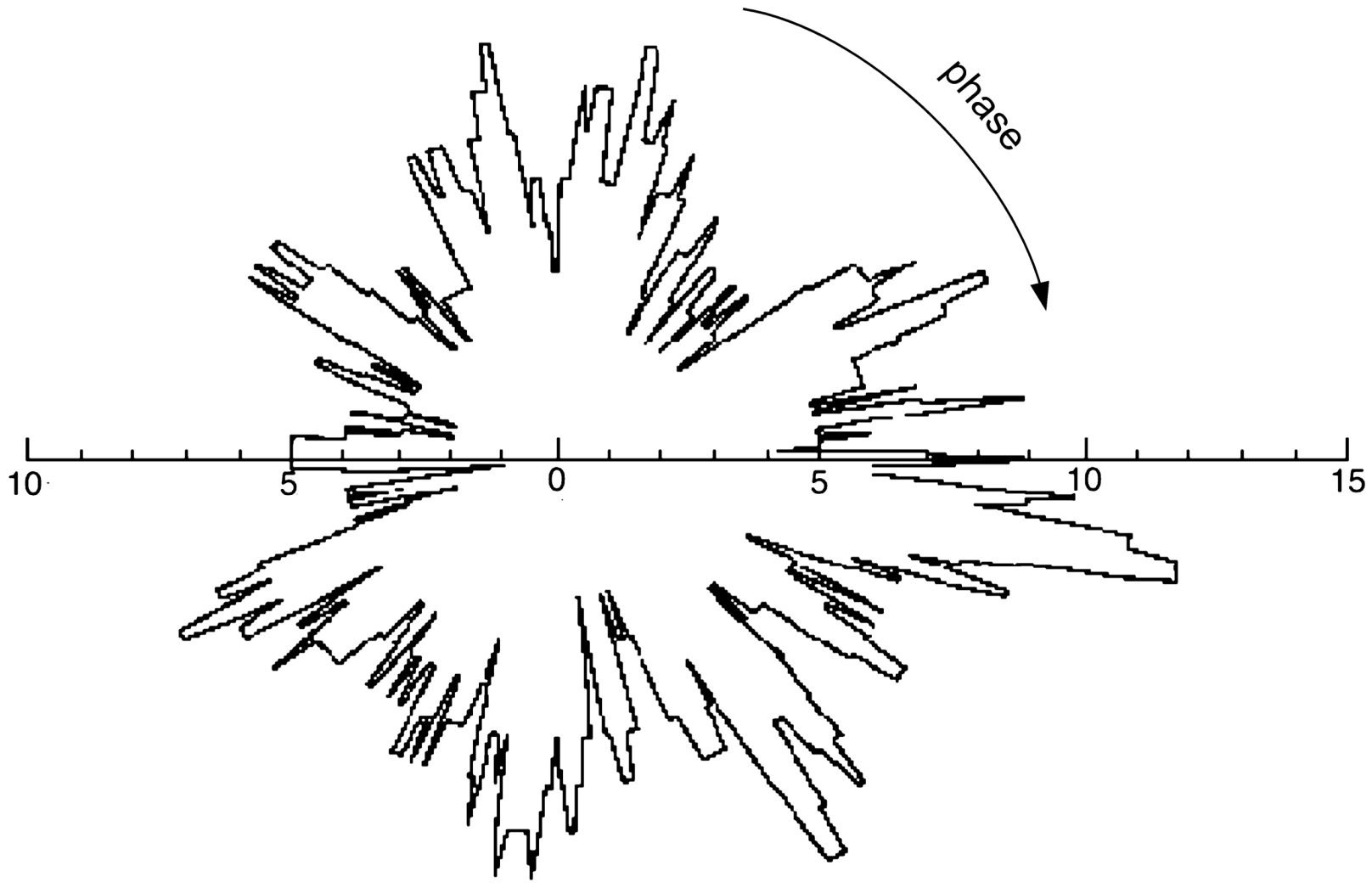

**Fig. 5**

The expected number of transits by a test planet with a period of 5 days that would have been observed in a ground-based light curve covering 617 h (= 25.7 d). The clockwise direction is the planet's phase at some time of reference, and the radial distance gives the number of transits that would have been observed at each phase. Achieving that the large majority of the potential phases would produce at least three transits required an observational coverage that was much longer than three times the orbital period. (Adapted from Deeg et al. 1998)

*Transit detection precision*

A basic version of the S/N of a single transit is given by the ratio

$$(S/N)_{tr} \approx \Delta F \,/\, \sigma_{lc}\,, \tag{9}$$

where $\Delta F$ is the fractional flux loss during a transit and $\sigma_{lc}$ is the fractional noise of the light curve on the timescale of the transit duration $TT$. This noise is composed of various sources, most notably photon noise from the target and the surrounding sky background , cosmic ray hits, CCD read noise, and flat-fielding or jitter noises (which arise from variations of the positions or shapes of stellar point spread functions on detectors whose sensitivity is not uniform). For ground-based surveys, we also have to add variations from atmospheric transparency and scintillation noise. Figure 6 shows the scatter over 1 h timescales from the most precise space-based survey, Kepler, and from NGTS, one of the leading ground-based surveys.

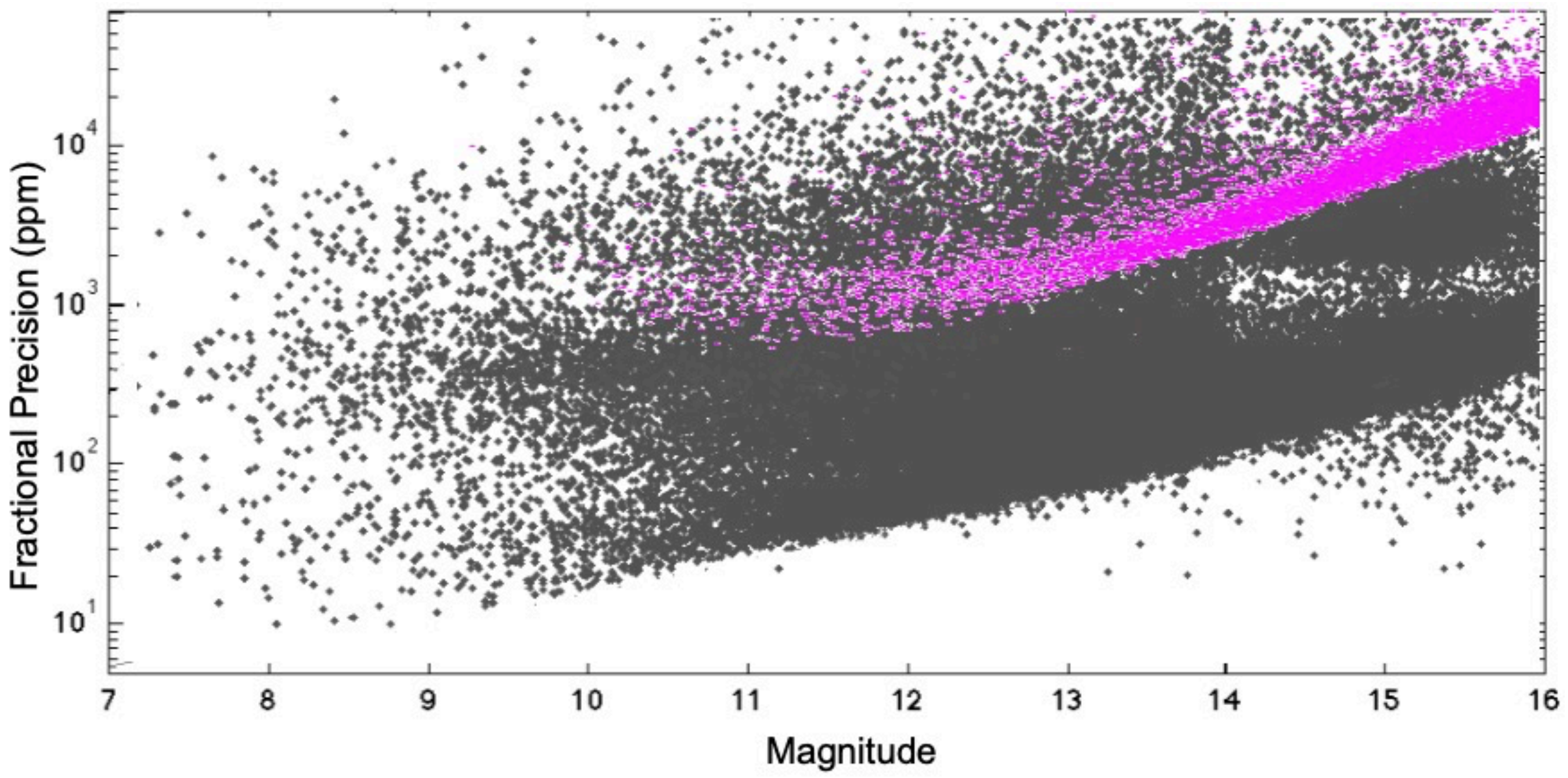

**Fig. 6**

Comparison of the precision between Kepler (gray points) and NGTS (violet points) photometry. Given is the light curves' rms scatter on a 1h timescale. The magnitudes for NGTS are in the I-band; for Kepler they are in its own system. The Kepler noises are from long-cadence (0.5h cycle time) curves of Q1 targets (Jenkins et al. 2010a) and scaled by $\sqrt{1/2}$ toward the 1h time-scale. The NGTS data are from 695 h of monitoring with a 12 s cadence, rebinned to exposure times of 1 h, from Wheatley et al. (2018). The difference between NGTS and Kepler precision would be reduced by a factor of $\sqrt{950/200} \approx 2.2$ if the different telescope aperture sizes are taken into account. The precision for the brightest NGTS targets is limited by scintillation noise, which is independent of the targets' brightness

Early estimates for planet detection yields assumed commonly a white-noise scaling from the point-to-point scatter of an observed light curve to the usually much longer duration of a transit. In practice, red or correlated noises degrade the precision of nearly all photometric time series over longer timescales, as was first shown by Pont et al. (2006), based on data from the OGLE-III transit survey. Only the space-based data from the Kepler mission uphold a white-noise scaling from their acquisition cycle of 30 min to a transit-like duration of 6 h (Jenkins et al. 2010a ; see Gilliland et al. 2011 for more details on Kepler's noise properties). The CoRoT mission, in contrary to Kepler on a low Earth orbit, produced light curves that on timescales of 2 h were already about twice as noisy as would be the result of a white-noise scaling from their acquisition cycle of 8.5 min (Aigrain et al. 2009). At least as strongly affected are ground-based surveys, with the principal culprit being the nightly air mass variation, which is on a similar timescale as the duration of most transits. Correlated noises have been the principal source for the early overestimations of detection yields. In the case of SuperWASP, recognizing their influence led to a revision of detection yields and to an increase in temporal coverage early in its operational phase (Smith et al. 2006). For surveys that attempt to detect shallow transits, brightness variations due to the sample stars' activity might also be of concern. The demonstration that this variability does *not* prevent the detection of terrestrial planets of solar-like stars (Jenkins 2002) was an important advance during the development of the Kepler mission. Aigrain et al. (2004) found then that K stars are the most promising targets for transit surveys, while a survey's performance drops significantly for stars earlier than G and younger than 2.0 Gyr. For a quantitative discussion of the factors that influence the yield of transit surveys, we refer to Beatty and Gaudi (2008).

Algorithms to dampen red noises and other systematic effects have been developed to either "clean" directly a light curve from their influences or as part of a detection algorithm, thereby increasing its sensitivity towards transit-like features. Examples are the pre-whitening employed in the Kepler pipeline (Jenkins et al. 2010b) , the cleaning of CoRoT light curves (Guterman et al. 2015), or the widely used SYSREM (Tamuz et al. 2005) and TFA (Kovács et al. 2005) algorithms.

*Brightness of the sample: Rejection of false positives and characterization of the planet catch*

As mentioned, a large number of transit surveys were initiated in the first years of the twenty-first century, after the discovery of the first transiting planets. These early efforts were aimed about equally at deep surveys of small fields using larger (1m and more) telescopes and at shallow surveys with small instruments having wide fields of view. The surveys with larger telescopes, including early projects with Hubble Space Telescope (Gilliland et al. 2000, on the 47 Tucanae globular cluster, and the SWEEPS survey by Sahu et al. 2006), were met however with limited success, with the most productive one becoming the OGLE-III (Udalski 2003) survey using a dedicated 1.3m telescope. Besides the difficulties to get access to the required large facilities to perform a deep survey over a sufficiently long time, a major drawback of such surveys is the faintness of the sample. RV verifications or further observational refinements of their transit detections are either impossible, or if possible at all, they likely require the largest existing telescope facilities.

For example, from the SWEEPS survey that targeted the Sagittarius I window of the Galactic bulge with the Hubble Space Telescope, Sahu et al. (2006) report the detection of transits on 16 targets. Their faintness of V = 18.8–26.2 as well as crowding permitted however only for the two of them (SWEEPS-04 and SWEEPS-11) a confirmation as planets, based on RVs taken with the 8 m VLT. All other SWEEPS detections have remained in candidate status until the present. We also note the comparatively small impact (relative to brighter targets) of the very large number of planets found on the fainter end of the Kepler mission's sample (Rowe et al. 2014 with 815 planets, Morton et al. 2016 with 1284 planets). These planets count only with probabilistic validations, and their principal usefulness are statistical studies on planet abundances across their known parameters (radius, period, central star type, planet multiplicity). The brightness of a target sample is therefore a very valuable parameter toward the science return of a transit survey!

The most common follow-up observations of transit detections are RV measurements, which do not only prove (or disprove) a planet's existence beyond reasonable doubt but also greatly improve our knowledge about them, providing masses, orbital eccentricity, and, occasionally, also the detection of further non-transiting planets in the same system. In practice, from the RV follow-up of numerous candidates for the Kepler, K2, and CoRoT missions, we found that a magnitude of ≈14.5 is a soft limit for their routinary follow-up. This is due to that brightness being near the limit for RV measurements at several relatively well-accessible midsized telescopes with appropriate instrumentation (e.g., the FIES instrument on the 2.5 m Nordic Optical Telescope or the HARPS instruments on the 3.6 m Telescopio Nazionale Galileo (TNG) and on the ESO 3.6 m telescope).

For transiting systems of bright central stars, a host of further possibilities to examine these systems opens up – such as observation of the Rossiter-McLaughlin effect, transit spectroscopy, secondary eclipse measurements, or the detection of phase curves (see this handbook's part on Exoplanet Characterization). For this reason – increased knowledge about the discovered systems – both the current TESS and the upcoming PLATO missions focus on samples that are brighter than those of Kepler and CoRoT, while ground-based surveys continue with their efforts to find transiting planets principally on bright or on special types of target stars.

# Transit Detection in Light Curves

Efficient recognition of transit-like features in light curves is a central part of any transit detection experiment. This task is usually performed in two steps. In the first one, detection statistical values that describe the likelihood of a light curve to contain a transit-like event are assigned. These values might also be expressed as a function of a candidate planet's size, period, and further parameters. In the second step, these statistical values are evaluated, and those candidates that deserve closer investigation are extracted. We reproduce here a description of this step in the Kepler pipeline, from Jenkins et al. (*2010b*): "Light curves whose maximum folded detection statistic exceeds 7.1$\sigma$ are designated Threshold Crossing Events (TCEs) and subjected to a suite of diagnostic tests in Data Validation (DV) to fit a planetary model to the data and to establish or break confidence in the planetary nature of the transit-like events." The threshold value for the extraction of candidates needs to be chosen with care, as it must provide a balance between the number of false positives – which increases to unmanageable levels if the threshold is too low – and the risk to miss detections of true planets if the threshold is too high.

As representative transit detection methods and algorithms, we mention here the early work on matched-filter detection algorithms by Jenkins et al. (*1996*) , which provided the basics for the transit detection of the early TEP observing project as well as for the Kepler mission, the widely used box least-squares (BLS) algorithm (Kovács et al. *2002*) with derivatives (e.g., Collier Cameron et al. *2006,* TLS: Hippke & Heller 2019), or algorithms using wavelets (e.g., Régulo et al. *2007*).

For the second step of a detection procedure, the evaluation of a transit candidate as a planet-like event, usually a more detailed modelling (or fitting) of the light curve of the presumed transit is performed. The Mandel and Agol (*2002*) algorithm and the analytical eclipsing formulae by Giménez (*2006*) are widely used basic transit modellers that have also been integrated into several transit fitting packages. For an overview of such tools, we refer to Chap. 76, "Tools for Transit and Radial Velocity Modeling and Analysis".

Table 1 Selected Transit Surveys

| Instrument | Years of operation | Configuration | *fov* $_{single}$ (deg$^2$) | *fov* $_{instr}$ (deg$^2$) | $N_{planet}$ | $R_{05}$,$R_{50}$,$R_{95}$ ($R_{jup}$) | $m_{05}$, $m_{50}$, $m_{95}$ (Vmag) |
|---|---|---|---|---|---|---|---|
| | | **Ground-based surveys** | | | | | |
| OGLE-III | 2001-2009 | Single telescope of 1300 mm Ø | 0.34 | | 8 | 1.08, 1.25, 1.61 | 14.0, 15.0, 15.8 [b] |
| TrES | 2003-2010 | 3 sites, each with one telescope of 100 mm Ø | 36 | | 5 | 1.10, 1.21, 1.71 | 11.4, 11.8, 13.7 |
| XO | 2003-2014 | Single site (3 sites in 2012-14) with common mount for 2 cams. | 49 | 300 | 9 | 0.97, 1.21, 2.07 | 9.8, 11.1, 12.2 |
| HATnet | since 2003 | 2 sites , one (FLWO) with 5 cams and one (Mauna Kea) with 2 cams, all individually mounted, with 110 mm Ø | 64 | | 67 | 0.87, 1.27, 1.89 | 9.7, 11.9, 13.8 |
| HATSouth | since 2009 | 3 sites, each with 2 mounts that each hold 4 cameras with 180 mm Ø, covering adjacent fields in 2x2 pattern | 17 | 67 | 73 | 0.69, 1.19, 1.68 | 12.0, 13.6, 16.2 |
| SuperWASP WASP South | 2004 - 2015 | 2 sites, each with single-mount array of 8 cams of 111 mm Ø covering adjacent fields | 60 | 480 | 161 | 0.85, 1.22, 1.62 | 9.8, 11.6, 13.0 |
| KELT | 2005 - 2020 | 2 sites, each with single camera of 42 mm Ø | 676 | | 21 | 1.2, 1.57, 1.9 | 7.6,10.0, 11.4 |
| NGTS | since 2016 | Single site with 12 telescopes of 200 mm Ø on independent mounts | 8 | 96 | 20 | 0.4, 1.21, 1.5 | 12.3, 14.0, 15.5 |
| ARGUS | since 2023 (Pathfinder) | Single site with 900 telescopes (Pathfinder: 38 telescopes) of 203mm aperture on common mount | 9.04 | 7916 | | | |
| | | **Space-based surveys** | | | | | |
| SWEEPS | 2006 | Hubble Space telescope (2.5 m Ø) with ACS camera | 0.003 | | 2 | 0.97 [c] | 19.3 [c] |
| CoRoT | 2007-2012 | Space telescope of 270 mm Ø | 3.5 / 1.7 [a] | | 35 | 0.3, 1.05, 1.47 | 11.9, 14.9, 15.8 |
| Kepler | 2008-2013 | Space telescope of 950 mm Ø | 107/102 [d] | | 2729 | 0.083, 0.19, 0.58 | 12.1, 14.8, 16.2 |
| K2 | 2014-2018 | Same as Kepler | 97 / 92 [d] | | 547 | 0.092, 0.20, 0.88 | 10.2, 12.9, 16.1 |
| TESS | since 2018 | 4 cams of 105 mm Ø, each covering adjacent areas | 576 | 2300 | 412 [e] | 0.105, 0.29, 1.44 | 8.3, 11.3, 15.6 |
| PLATO | launch end of 2026 | 24 cams of 100 mm Ø in 4 groups, with 6 co-aligned cams in each group. Partial overlap of all 24 units. 2 further 'fast cams' with color filter and short cycle time | 1100 | 2232 | | | |

Notes

[a] Original *fov* and *fov* after March 2009, due to on-board failure

[b] I band magnitudes

[c] Average of the two confirmed planets

[d] *fov* values before / after Jan 2010 (Kepler) and before / after July 2016 (K2), due to detector failures

[e] Number from 'TESS as discovery facility' OR 'TOI in planet name' (only discoveries by transits or TTVs)

# Transit Surveys: Past, Current, and Future Projects

The Extrasolar Planets Encyclopaedia (at http://www.exoplanet.eu/research/) lists currently about 50 planet search projects that indicate "transits" as a principal observing method. These projects include finished ones, currently operating ones, projects that are in various preparation stages, as well as projects or proposals that have never moved beyond some design phase. It includes also some projects that aren't dedicated to the discovery but to the follow-up of transiting planets, such as ESA's CHEOPS space mission. In Table 1 and in the following notes, we provide an overview over a selection of well-known transit detection surveys. For most of them, this handbook contains also dedicated chapters, listed in the Cross-References. The columns of Table 1 have the following meaning:

- years: The years of operation
- config: Instrument configuration, with the aperture diameters of the individual optical units (cam = camera).
- $fov_{single}$: The sky area in deg$^2$ covered by a single optical unit of the detection experiment.
- $fov_{instr}$: The sky area in deg$^2$ that is covered simultaneously be the experiment at a single site in its usual operating mode. Only given if there are multiple optical units at a site.
- *Npl*: Count of detected planets. Only discoveries based on transits or transit timing variations are included.
- $R_{05}$, $R_{50}$, $R_{95}$: 5th, 50th (median), and 95th percentiles of the radii of the detected planets. If $Npl \leq 20$, the smallest and largest planets are given.
- $m_{05}$, $m_{50}$, $m_{95}$: Similar to the previous but indicating the V-mag brightness of the detected systems.

Planet counts, radii, and magnitudes are from the NASA Exoplanet Archive, obtained in December 2023. Below, some notes are provided for the transit surveys listed in Table 1.

*OGLE-III*

The Optical Gravitational Lensing Experiment has been implemented in four phases, with the fourth one (Udalski et al. 2015) operational at present (Fall 2023). OGLE is dedicated to the detection of substellar objects from microlensing, except during its third phase (OGLE-III, Udalski *2003*), when the observing procedure of the 1.3m OGLE telescope at Las Campanas Observatory was modified to enable the detection of transits. OGLE-TR-56 was the first planet discovered in a transit search, with a posterior verification from RV follow-up (Konacki et al. *2003*).

*TrES*

The "Transatlantic Exoplanet Survey" was the first project with instruments that were specifically designed and dedicated for transit surveying. Its first telescope, originally named STARE, was used in the 1999 discovery of the transits of HD 209458b during tests at the High Altitude Observatory at Boulder, Colorado. In 2001, it was relocated to Teide Observatory, Tenerife, where a systematic transit search began. Since 2003, the project operated under the TrES name, after the merger with two other projects using similar instrumentation, namely, PSST at Lowell Observatory and the Sleuth Project at Palomar Observatory (O'Donovan *2008*). The principal success of TrES was the detection of the first transiting planets orbiting bright stars (TrES 1, Alonso et al. *2004a*; TrES, 2 O'Donovan et al. *2006*) by a dedicated survey. TrES was decomissioned in 2010.

*XO*
This survey started in 2003 at a single site, with a second phase observing from three sites from 2012 to 2014. The CCDs were red in time-delayed integration (TDI): pixels are red continuously, while stars move along the columns on the detector, owing to a slewing motion of the telescope. This setup enlarges the effective field of view and resulted in stripes of 7°× 43° that were acquired during each single exposure.

*HAT*
This denominator (Hungarian-made Automated Telescope) encompasses two surveys: For one, HATnet operates since 2003 seven CCD cameras with 110 mm apertures on individual mounts, with five of them at Fred Lawrence Whipple Observatory at Mount Hopkins in Arizona and two at Mauna Kea Observatory in Hawaii. For another, HATSouth (Bakos et al. *2013*) is a network across three sites in the southern hemisphere that is able to track stars continuously over longer time spans. Since 2009, it operates at Las Campanas Observatory (Chile), at the High Energy Stereoscopic System site (Namibia), and at Siding Spring Observatory (Australia), with limited activities continuing at present (2023). Each of these sites contains two mounts, with each of them holding four Takahashi astrographs with individual apertures of 180 mm. The HAT consortium is also advancing the HATPI Project of an all-sky camera consisting of 63 optical units on a single mount.

*SuperWASP and WASP-South*
The WASP (Wide Angle Search for Planets; see Pollacco et al. *2006* for an instrument description; Smith et al. *2014* for a review.) surveys consisted of two instruments: SuperWASP, since 2004 at Roque de los Muchachos Observatory on the Canary Island of La Palma, and WASP-South, since 2006 at the South African Astronomical Observatory. A predecessor instrument, WASP0, was operated during the year 2000 on La Palma. SuperWASP was an array of 8 cameras covering $480^2$ of sky with each exposure; and WASP-South was a close copy of it; both instruments were decommissioned around 2018. With 161 planets that were primarily discovered by one of the WASP surveys, and 179 'WASP'-named planets (as of Oct. 2023), WASP has contributed to the detection of more planets than any other ground-based effort. Among these discoveries are several exoplanets (such as WASP-3b, 12b, 43b) that stand out for their excellent suitability for deeper characterization work, due to their short orbital period and/or large size.

*KELT*
The "Kilodegree Extremely Little Telescope" has to date been the most successful survey using very wide-field detectors (with a *fov* of 26°× 26°) , using commercial photographic optics of short focal length. KELT-North operated since 2005 from Winer Observatory, Arizona, and KELT-South since 2009 from Sutherland, South Africa; the KELT survey concluded in 2020. Both instruments used a CCD camera with an 80 mm/f1.8 Mamya lens. KELT has detected the hottest exoplanet currently known, KELT 9-b with an equilibrium temperature of 4050 K (Gaudi et al. 2017).

*NGTS*
The Next-Generation Transit Survey (Wheatley et al. *2013*, *2018*) is operated by a consortium of seven institutions from Chile, Germany, Switzerland, and the United Kingdom. After testing in La Palma and at Geneva Observatory, operations started in 2016 at ESO's Paranal Observatory. NGTS employs an automated array of twelve 20-centimeter f/2.8 telescopes on independent mounts, sensitive to orange to near-infrared wavelengths (600–900 nm). It is a successor project to WASP that achieves significantly better photometric precision (Fig. 7) but with a focus on late-type stars. Its first planet discovery was the most massive planet known to transit an M-dwarf (Bayliss et al. 2018), and it has also discovered two hot Saturn-sized planets, NGTS-4b (West et al. 2019) and NGTS-14Ab (Smith et al. 2021). The high precision of NGTS is also manifested in the rather faint average magnitude of 14 among the planets discovered by this survey (see Table 1). NGTS is also contributing to the follow-up of planet detections by other consortia, especially for the TESS space mission.

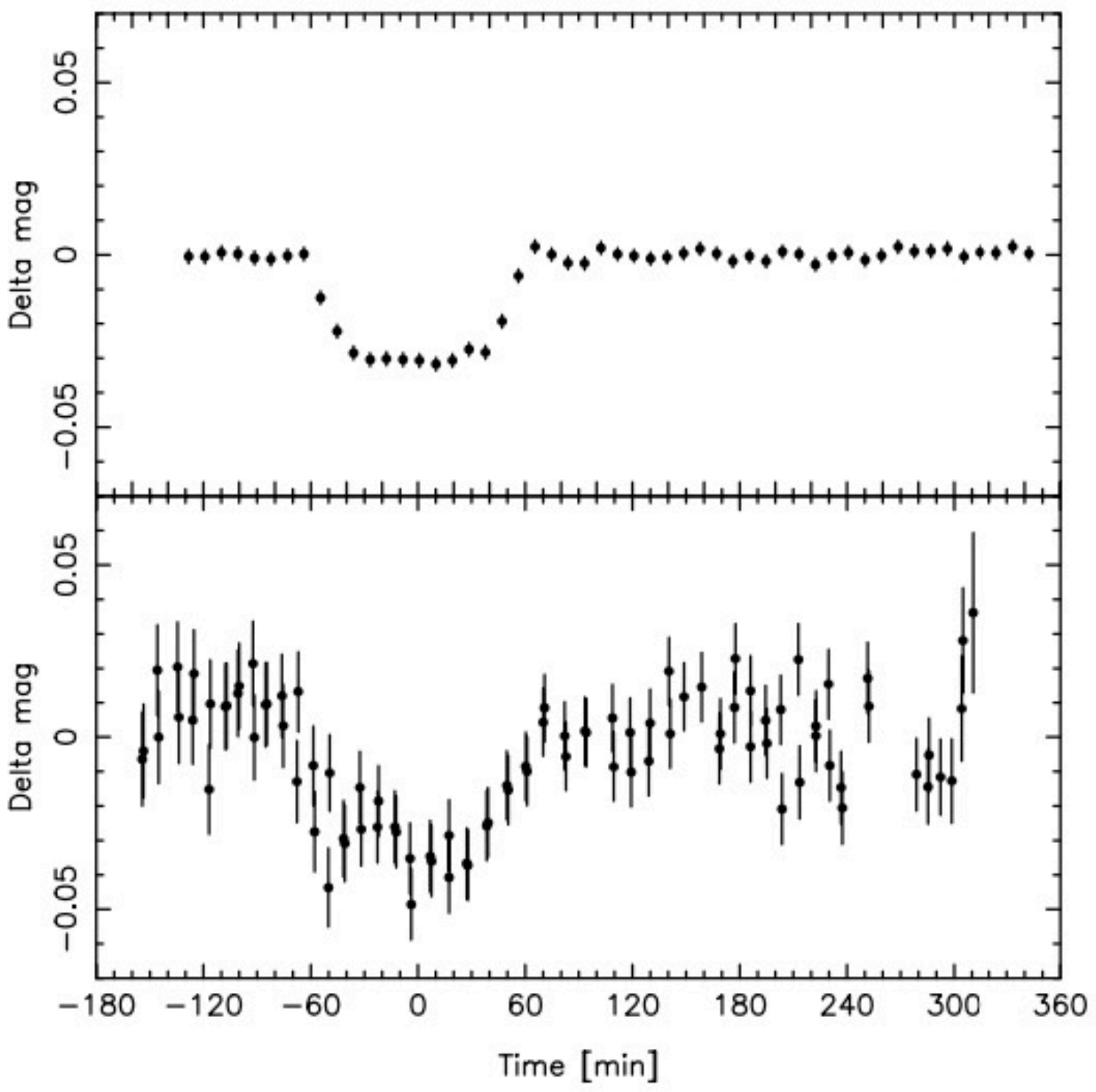

**Fig. 7**

Single transit observations of the Hot Jupiter WASP- 4b with one NGTS telescope unit (top) and WASP (bottom). (From Wheatley et al. 2018, reproduced with permission)

*ARGUS*

ARGUS, led by the University of North Carolina at Chapel Hill (Law et al. 2022a), is an ambitious survey program for the Northern Hemisphere. It features an array of 900 telescopes with 20 cm apertures and 61Mpix cameras covering the entire visible sky, with all telescopes placed on a single mount. ARGUS contemplates multiple objectives in time-domain astronomy, among them the study of exoplanets, stellar science, the exploration of general transient astronomical phenomena, and solar system science. ARGUS will have a one-minute temporal resolution, with exposures of one minute and one second, and it will track stars for approximately 15 minutes before resetting the entire mount. Two instrument designs are currently under consideration (Law et al. 2022a, Law et al. 2022 b). The ARGUS pathfinder prototype, a scaled-down version with 38 telescopes, has been in the commissioning since early 2023 at the PARI observatory in North Carolina.

*CoRoT*

Named after "Convection, Rotation, and Transits," this was the first space mission dedicated to exoplanets. Launched in December 2006 by the French space agency CNES and partners into a low polar orbit for a survey lasting initially 4 years, it surveyed 163,665 targets distributed over 26 stellar fields in two opposite regions in the galactic plane, with survey coverages lasting between 21 and 152 days (Deleuil et al. 2018). In May 2009, its first data processing unit failed, and CoRoT's *fov* was reduced to half, while the failure of the other unit in Nov. 2012 caused the end of the mission. Its most emblematic discovery was CoRoT-7b, the first transiting terrestrial planet (Léger et al. 2009)

*Kepler*

This NASA mission was launched in 2009 into an Earth-trailing orbit, for a mission of 4 years to survey a single field of 170,000 stars, principally for the presence of Earth-sized planets. Kepler has discovered the majority of currently known exoplanets, with discoveries that have revolutionized the field of exoplanets. In contrary to the planets found by any other transit survey, only a small fraction (3%) of Kepler planets are Jupiter-sized ($\geq 0.9R_{\mathrm{jup}}$), while the vast majority are Earth- or super-Earth-sized ones. Science operations under the "Kepler" denomination ended in May 2013 when two of the spacecraft's reaction wheels failed and its pointing become unreliable.

*K2*

In March 2014, the Kepler spacecraft was returned into service under the K2 name. Its observing mode was adapted to the reduced number of reaction wheels, surveying fields near the ecliptic plane for about 80 days each (Howell et al. 2014). Planets found by K2 have a rather similar size distribution to Kepler, albeit with a somewhat larger fraction of giant planets (5% are larger than $0.9\mathrm{R}_{\mathrm{jup}}$). The spacecraft was deactivated in October 2018, when it ran out of fuel.

*TESS*

The "Transiting Exoplanets Survey Satellite" by NASA aims to scan nearly the entire sky for transits across relatively bright stars (Ricker et al. 2015). Most areas are being covered by pointings lasting 28 days. The spacecraft harbors four wide-field telescopes that cover jointly a stripe of the sky of 24° by 96°. TESS was launched in April 2018 into an elliptical orbit with a 13.7-day period in a 2:1 resonance with the Moon's orbit, for a mission with an initial duration of 2 years. Given the success of that mission, it has been extended already twice, with operations of its second extended mission concluding in October 2024.

*PLATO*

This ESA mission , named after "PLAnetary Transits and Oscillation of stars," is scheduled to be launched at the end of 2026 into an orbit around the L2 point, to perform during at least 4 years a survey of several large sky areas (Rauer et al. 2014), with an initial field for a 2-year long survey that will be in the southern hemisphere. The mission's core sample are 15,000 stars of $8 \leq mV \leq 11$, while a secondary "statistical" sample includes 245,000 targets up to $mV \approx 16$. PLATO will have four groups of detectors, each with six cameras that all point to the same *fov*. Between the groups there is a partial overlap due to which areas near the center of the common *fov* will be covered by all 24 cameras, while outer zones will be covered by 6 or 12 cameras only. Two additional "fast" cameras with rapid cycle times and color filters will survey the brightest stars of 4–8 $mV$.

## Surveys for Planets of Low-Mass Stars

Several surveys, which are not listed in Table 1, have been designed specifically for the detection of planets around low-mass stars, and in particular, M-stars. Given the difficulties to detect planets in the habitable zone of solar -like stars, planet searches around such stars provide an alternative path for the detection of potentially habitable planets (e.g., Scalo et al. 2007). Their small size permits that terrestrial planets produce transits that are deep enough to be observable from moderate ground-based instruments. Also, the habitable zone around these stars corresponds to orbital periods of a few days to weeks, making habitable planets' transits shorter, more frequent, and hence easier to detect than for solar-type stars. Disadvantages of low-mass stars as targets are however a flux variability that is exhibited by most of them and the sparsity of such stars with sufficient apparent brightness. As a consequence, these detection projects are not performed as wide-field surveys but as searches that point to selected target stars, which are covered sequentially. As such, these projects cover relatively few targets and have only a small planet catch but may provide discoveries of large impact toward our knowledge of potentially habitable planets.

*MEarth*

This project operates since 2008, consisting of eight 40 cm telecopes at Mount Hopkins, Arizona, and since 2014, of a similar setup at Cerro Tololo, Chile (Berta et al. *2012*). MEarth has discovered several small planets, among them LHS1140b, a planet of 1.4 $R_{\text{Earth}}$ in the habitable zone of an M-dwarf at a distance of 10.5 parsec (Dittmann et al. *2017*).

*TRAPPIST/SPECULOOS*

The " TRAnsiting Planets and PlanetesImals Small Telescope" survey consists of two 60 cm robotic telescopes, one operating since 2010 at ESO's La Silla Observatory, Chile, and one since 2016 at Oukaimeden Observatory, Morocco. It has the dual objective of transit detection and the study of comets and other small bodies in the Solar System (Jehin et al. *2014*). Its outstanding discovery has been the TRAPPIST-1 system of seven planets, with some of them in the habitable zone, around an ultracool M8 dwarf at a distance of 12 parsec (Gillon et al. *2017*). TRAPPIST is also a prototype of the SPECULOOS (Search for habitable Planets EClipsing ULtra-cOOl Stars) project, whose first phase consists of four 1m robotic telescopes at ESO's Paranal Observatory (Delrez et al. 2018), and two additional ones at San Pedro Martir (Demory et al. 2020), and Teide Observatory (Burdanov et al. 2022), respectively.

# Conclusion

In the year 2003, K. Horne predicted the success of transit surveys in a paper entitled "Hot Jupiters Galore." It took longer than expected to get to that point and required the understanding and resolution of several subtle issues affecting these surveys, but today the paper's title has become reality, and the discovery of transiting planets is commonplace. This applies not only to Hot Jupiters but also to planets across the entire size regime and has been a consequence of the continued refinement of observing techniques and of the development of new instruments, both ground and space based.

At the time of writing, the transit method is expected to remain the largest contributor toward the discovery of new planets and planet systems, with several ambitious ground- and space-based searches under way. Planet systems found in transit searches will also continue to provide the motivation for the continued development of instruments and observing techniques, which take advantage of the opportunities for deeper insights that transiting systems are offering. In that sense, systems found by transit surveys will continue as a basic nutrient of the field of exoplanet science.

For further reading about transits as a tool to detect and characterize exoplanets, we refer to the reviews by Winn (*2010*) and by Cameron (*2016*) and to a book dedicated to transiting exoplanets by Haswell (*2010*).

## Cross-References to chapters in the Handbook of Exoplanets

- Characterization of Exoplanets: Secondary Eclipses
- CoRoT: The First Space-Based Transit Survey to Explore the Close-in Planet Population
- Discovery of the First Transiting Planets
- Exoplanet Phase Curves: Observations and Theory
- KELT: The Kilodegree Extremely Little Telescope, a Survey for Exoplanets Transiting Bright, Hot Stars
- Microlensing Surveys for Exoplanet Research (OGLE Survey Perspective)
- Planet Occurrence: Doppler and Transit Surveys
- Populations of Extrasolar Giant Planets from Transit and Radial Velocity Surveys
- Prehistory of Transit Searches
- Radial Velocities as an Exoplanet Discovery Method
- Small Telescope Exoplanet Transit Surveys: XO
- Space Missions for Exoplanet Science: Kepler/K2
- Space Missions for Exoplanet Science: PLATO
- SPECULOOS Exoplanet Search and Its Prototype on TRAPPIST
- The HATNet and HATSouth Exoplanet Surveys
- The Rossiter-McLaughlin Effect in Exoplanet Research
- Tools for Transit and Radial Velocity Modeling and Analysis
- Transit-Timing and Duration Variations for the Discovery and Characterization of Exoplanets

**Acknowledgements**

For the second edition, the authors acknowledge support from the Spanish Research Agency of the Ministry of Science and Innovation (AEI-MICINN) under grant 'Contribution of the IAC to the PLATO Space Mission' with reference PID2019-107061GB-C66, DOI: 10.13039/501100011033. This contribution has benefited from the use of the NASA Exoplanet Archive and the Extrasolar Planets Encyclopaedia, and the authors acknowledge the people behind these tools.